\documentclass[10pt]{article}
\usepackage{epsf}
\usepackage{latexsym}

\newcommand{\nix}[1]{}
\begin{document}
\title{\large \bf Optical Spin Orientation
\\
under Inter- and Intra-Subband Transitions in QWs}
\author{S.A.~Tarasenko$^1$, E.L.~Ivchenko$^1$, V.V.~Bel'kov$^1$,
S.D.~Ganichev$^{2}$, D.~Schowalter$^{2}$,  Petra~Schneider$^2$, \\M.~Sollinger$^2$,
W.~Prettl$^2$, V.M.~Ustinov$^1$,  A.E.~Zhukov$^1$,  and L.~E.~Vorobjev$^3$}

\date{\normalsize
$^1$ A.F.~Ioffe Physico-Technical Institute, 194021 St. Petersburg,
Russia
\\$^2$ Fakult\"{a}t f\"{u}r Physik, Universit\"{a}t Regensburg, D-93040 Regensburg,
Germany\\
$^3$ St. Petersburg State Technical University, 195251 St.~Petersburg, Russia\\
}
\thispagestyle{empty}
\onecolumn
\maketitle
{\bf Abstract:}

It is shown that absorption of circularly polarized infrared radiation
achieved by inter-subband and intra-subband (Drude-like) transitions results
in a monopolar spin orientation of free carriers. The monopolar spin
polarization in zinc-blende-based quantum wells (QWs)
is demonstrated
by the observation of the  spin-galvanic  and
circular photogalvanic effects. It is
shown that  monopolar spin orientation
in n-type QWs becomes possible if
an admixture of
valence band states to the conduction band wave function and the
spin-orbit splitting of the valence band are taken into account.
\section{{Introduction}}

Absorption of circularly polarized light in semiconductors may
result in spin polarization of photoexcited carriers. This
phenomenon of optical orientation is well known for interband
transitions in  semiconductors~\cite{Meier}.
In this paper we show that the absorption of terahertz radiation with
photon energies much less than  the energy gap also leads to spin orientation
of free carriers. This optical orientation may be referred to as
`monopolar' because only one type of carriers, electrons or holes, is
excited.

We present theoretical and experimental results on monopolar optical
orientation of a  two-dimensional electron gas or hole gas. Both direct
inter-subband and indirect (Drude-like) intra-subband transitions induced
by circularly polarized radiation are considered for $n$- and
$p$-type
QWs based on zinc-blende-structure semiconductors. A transfer of the photon
angular momentum to the electron spin is linked to the spin-orbit
interaction. Therefore, monopolar optical orientation of electrons can be
obtained if an admixture of the
$\Gamma_7$ and $\Gamma_8$ valence band states
to the conduction band wave functions is taken into account.

Generally the expression for the generation rate of electron spin
polarization due to
optical excitation can be written as
\begin{equation}\label{spingen}
\dot{S}=s (\eta I/ \hbar \omega) P_{circ} \:,
\end{equation}
where $s$  is the average electron spin generated per one absorbed photon
of circularly polarized radiation, $\eta$ is the fraction of the energy flux
absorbed in the QW, $I$ is the light intensity, $\hbar\omega$ is the
photon energy, and $P_{circ}$ is the degree of circular polarization.
While optical orientation at interband excitation has been
widely  studied, it is not obvious that  inter-subband and intra-subband
transitions can result in a spin polarization. Below we show that for both
kinds of transitions within one band - valence or conduction band - absorption
of circularly polarized light leads to  spin polarization which has not
been considered previously.

\section{{Experimental technique}}
The experiments have been carried out on MBE (001)-grown
$n$-GaAs/AlGaAs QW of 7~nm width, $n$-GaAs/AlGaAs single heterojunction,
 In$_{0.2}$Ga$_{0.8}$As
QWs of 7.6~nm width, and on $p$-type GaAs/AlGaAs QWs of 20~nm width
MOCVD grown on  (001)-miscut substrates as well as (113)-MBE-grown
samples with QWs of various widths between 7 and 15~nm. Samples with free carrier
densities of about $2\cdot 10^{11}$ cm$^{-2}$ were studied in the
temperature range
from liquid helium to room temperature. A pair of ohmic contacts was
centered on opposite sample edges along the direction $x \parallel
[1\bar{1}0]$ (see inset in Fig.~1).

A high power pulsed mid-infrared (MIR) TEA-CO$_2$ laser and a far-infrared
(FIR) NH$_3$ laser were  used as radiation sources delivering 100\,ns
pulses with radiation power $P$ up to 100\,kW. Several lines of the CO$_2$
laser between 9.2\,$\mu$m and 10.6\,$\mu$m and of the NH$_3$-laser~\cite{PhysicaB99}
between $\lambda$ = 76\,$\mu$m and 280\,$\mu$m were chosen  for
excitation in the MIR and FIR range, respectively. In $n$-type samples the
MIR radiation induces direct optical transitions between the first and the
second subband of QWs. Direct transitions between heavy-hole and light-hole
subbands have been achieved in $p$-type samples applying FIR radiation in
the spectral range between 76\,$\mu$m and 148\,$\mu$m. In both $n$- and $p$-type
samples application of FIR radiation with photon energies less than the
separation of subbands leads to absorption caused by indirect optical transitions
in the lowest subband (Drude absorption).

The laser light polarization was modified from
linear to circular using a Fresnel rhombus and  quartz $\lambda/4$ plates
for MIR and
FIR radiation, respectively. The
helicity of the incident light was varied
according to $P_{circ} = \sin{2
\varphi}$ where $\varphi$ is the angle between the initial
plane of linear polarization and the optical axis of the
polarizer. Spin polarization has been investigated making use of the
circular photogalvanic effect (CPGE)~\cite{PRL01} and the spin-galvanic
effect~\cite{Nature02}.
For investigation of spin-galvanic effect
an in-plane magnetic
field $B$ up to 3~T has been applied as shown in the inset
of Fig.~1.
The samples were placed in a magneto-optical cryostat with a
split-coil superconducting magnet.
The current $j$ generated by polarized light in the unbiased structures
was measured via the voltage drop across a 50~$\Omega$ load resistor in a
closed circuit configuration. The voltage was recorded with a storage
oscilloscope. The measured current pulses of 100~ns
duration reproduce the temporal structure of the laser pulses.
\section{{Direct inter-subband transitions}}
With illumination of  (001)-oriented
$n$-type GaAs and InAs QWs  at oblique incidence of
MIR radiation of the CO$_2$ laser
a current signal proportional to the helicity
$P_{circ}$ has been observed indicating the circular photogalvanic
effect~\cite{PRL01}.
At normal incidence of radiation, at which the CPGE effect vanishes,
the spin-galvanic
current~\cite{Nature02}
is also observed applying an in-plane
magnetic field. Both effects are due to spin orientation, therefore the
observation of the CPGE and the spin-galvanic effect clearly gives
evidence that the
absorption of infrared circularly polarized radiation results in spin orientation.
The wavelength
dependence of the spin-galvanic effect obtained between 9.2~$\mu$m and
10.6~$\mu$m repeats the spectral behaviour of direct inter-subband
absorption between the first and the second subbands measured
in transmission by Fourier spectroscopy in  multipath geometry.
This unambiguously demonstrates that in this case the
spin orientation of $n$-type QWs is obtained by
inter-subband transitions.

We would like to emphasize that
spin sensitive inter-subband transitions in
$n$-type QWs have been observed
at normal incidence when there is no component of the electric
field of the radiation normal to the plane of the QW.
Generally it is
believed that inter-subband transitions in $n$-type QWs
can only be excited by infrared
light polarized in the growth direction $z$ of the QWs~\cite{book}.
Furthermore such transitions are spin insensitive and, hence, do not lead to
optical orientation. Since the argument leading to these selection rules is
based on the effective mass approximation in a single band model, the
selection rules are not rigorous.

\begin{figure}[!h]
\begin{center}
\mbox{\epsfxsize = 6cm \epsfbox{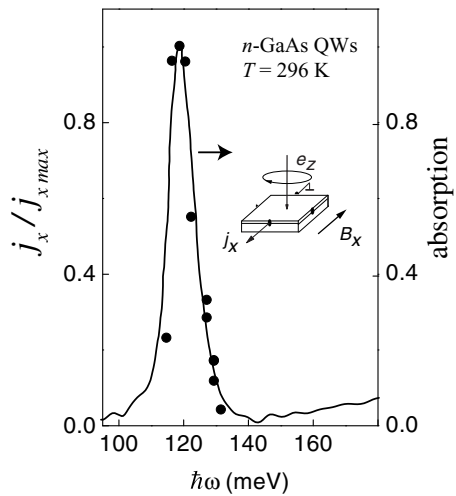}}
\end{center}
\caption{
Spectral dependence of the spin-galvanic current
(dots) caused by spin orientation due to direct optical transitions
between $e1$
and $e2$ conduction subbands.
The inset shows the geometry of the experiment. Optical excitation is at
normal incidence of right-handed cicularly polarized radiation.
For comparison the
absorption spectrum obtained from  transmission in a
multiple-reflection waveguide geometry shown by the full line.
Results
are plotted for (001)-grown GaAs QWs of 7~nm width at room temperature.}
\label{F1}
\end{figure}

In order to explain the observed spin orientation as well as the absorption
of light polarized
in the plane of the QW we show that
a ${\bf k} \cdot {\bf p}$ admixture of valence band states to the
conduction band wave functions has to be taken into account.
Calculations yield that
inter-subband absorption of circularly polarized light propagating
along $z$  induces only spin-flip
transitions resulting in  100\% optical orientation of
photoexcited carriers, $s=1$. In this geometry the fraction of the energy
flux absorbed in the QW by transitions from the first  subband $e1$ to the
second subband $e2$ has the form

\begin{equation}\label{eta_inter}
\eta = \frac{128 \alpha}{9 n_{\omega}} \,
\frac{\Delta^2_{so}(2E_g+\Delta_{so})^2 (E_2-E_1) E_1}
{E^2_g(E_g+\Delta_{so})^2(3E_g+2\Delta_{so})^2} \, \frac{\hbar^2 n_e}{m^* }
\,\delta(\hbar\omega - E_1 + E_2)  \: ,
\end{equation}

where $\alpha$ is the fine structure constant, $n_{\omega}$ is
the refraction index, $n_e$ is the 2D electron concentration, $m^* $
is the effective
mass, $\hbar\omega$ is  the photon energy,
$E_g$ is the energy gap,
$\Delta_{so}$ is the valence band spin-orbit splitting,
$E_1$ and $E_2$ are the energies of  size-quantization of the $e1$ and $e2$
subbands, respectively. The  $\delta$-function
describes the resonant behaviour of the inter-subband transitions.

The  helicity dependent current of the CPGE has also been observed in
$p$-type GaAs QWs due to transitions between heavy-hole (hh1) and light-hole (lh1)
subbands demonstrating  spin orientation of holes. The measurements were carried
out on  (001)-miscut  QWs and (113)-oriented QWs which, in contrast
to (001)-oriented samples and in accordance to the phenomenological theory of the CPGE,
yield a maximum of the CPGE at normal incidence of radiation. QWs with
various widths in the range from 7 to 20\,nm were investigated. For direct
inter-subband transitions photon energies between 35~meV and 8~meV of FIR
radiation corresponding to these QW widths were applied. Due to the
different effective masses of light and heavy holes the absorption
does not show narrow resonances. A spin-galvanic current in $p$-type QWs could not be
detected because of the very small in-plane $g$-factor for heavy
holes~\cite{gfactor} which makes the effect of the magnetic field
negligible~\cite{Nature02}. Optical orientation caused by heavy-hole to light-hole
absorption of circularly polarized radiation occurs for transitions at
in-plane wavevector ${\bf k} \neq 0$ due to the mixing of heavy-hole and light-hole subbands
\section{{Intra-subband transitions}}
In the far-infrared range, where the photon energy is not enough for direct
inter-subband transition in $n$- or even in $p$-type samples,
the absorption of light by free carriers occurs by indirect intra-subband  transitions
where momentum conservation law is satisfied due to
acoustic or optical phonons,  static defects etc.
In the case  of such free carrier absorption spin orientation is obtained as well.
This is proved by the observation of the CPGE in both $n$- and $p$-type samples
in response to FIR radiation.
For $n$-type samples also the
spin-galvanic effect caused by spin orientation was detected (Fig.~2).

\begin{figure}[!h]
\begin{center}
\mbox{\epsfxsize = 8cm \epsfbox{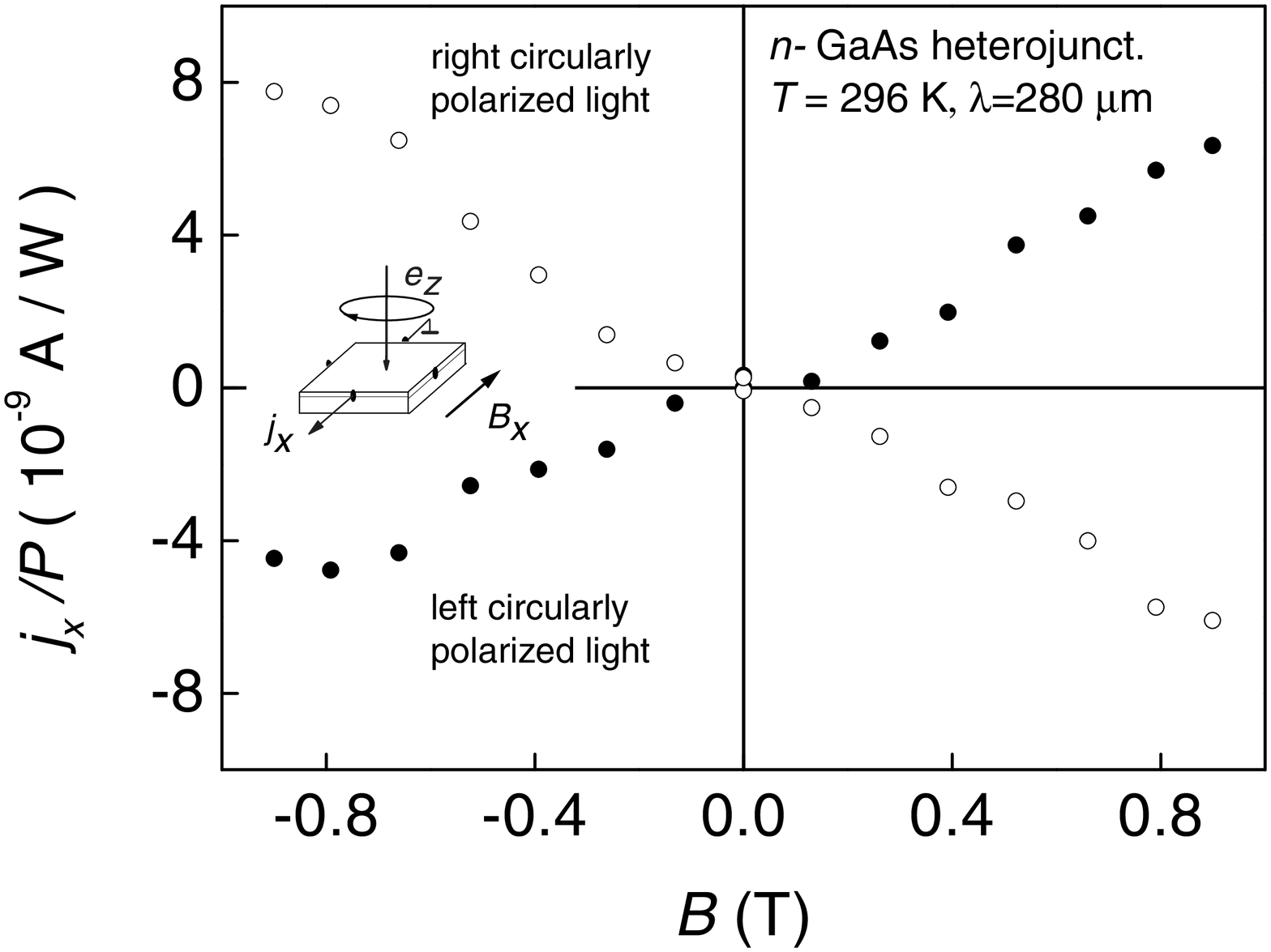}}
\end{center}
\caption{Magnetic field dependence of the spin-galvanic current
achieved by intra-subband transitions within $e1$
conduction subband by the excitation with the radiation of 280~$\mu$m
wavelength.
Results
are plotted for an (001)-grown GaAs single heterojunction at room temperature.}
\label{F2}
\end{figure}

The experiments were carried out on $n$-type (001)-oriented GaAs and InAs
QWs applying FIR radiation in the range from 76~$\mu$m to 280~$\mu$m
(16\,meV - 4.4\,meV which is much less than $E_2-E_1$ = 120~meV).
For $p$-type materials long wavelength  laser lines
from 280~$\mu$m to 496~$\mu$m
(4.4\,meV-2.5\,meV) were used to ensure free carrier intra-subband absorption for all QW
widths.
In this case
the photon energies are less than hh1-lh1 energy separation.

The intra-subband optical transitions in QWs involving
both the electron-photon interaction and momentum scattering are described by
second-order processes with virtual intermediate states. A
dominant contribution to the optical absorption is caused by
processes with intermediate states in the same subband.
This is the channel that determines the coefficient of intra-subband
absorption, $\eta$. However such transitions conserve the electronic spin and,
hence, do not lead to an optical orientation.

In order to obtain optical orientation due to intra-subband transitions we
considered virtual interband transitions
with intermediate states in the valence band. Taking into account the
spin-orbit splitting of the valence band and the
selection rules for
interband transitions, the absorption of circularly polarized light
gives rise to the spin orientation of electrons.

For this particular mechanism of monopolar optical orientation
one can derive the following expression for the
spin generated per one absorbed photon of
right-handed circularly polarized radiation

\begin{equation}\label{s_intra}
s \propto \frac{V^2_{cv}}{V^2_c}
\frac{\hbar\omega \, \Delta^2_{so}}{E_g(E_g+\Delta_{so})(3E_g+2\Delta_{so})} \:.
\end{equation}

Here $V_c$ and $V_{cv}$ are the intraband and interband matrix elements of
scattering, respectively. The prefactor in Eq.(\ref{s_intra}) depends on the
mechanism of momentum scattering. An estimation for typical GaAs/AlGaAs QWs
shows that $s \simeq 10^{-6}$ for acoustic phonon assisted indirect optical
transitions at the photon energy $\hbar \omega = 10$ meV.\\

In conclusion, our results demonstrate that in both $n$- and $p$-type QWs
monopolar spin orientation can be achieved applying radiation with photon
energies less than the fundamental energy gap. Spin orientation has been
observed by direct inter-subband transitions as well as by Drude-like
intra-subband absorption.
It is
shown that  monopolar spin orientation
in n-type QWs becomes possible if
an admixture of
valence band states to the conduction band wave function and the
spin-orbit splitting of the valence band are taken into account.
We emphasize that the spin generation rate under
monopolar optical orientation  depends strongly on the energy of spin-orbit splitting
of the valence band, $\Delta_{so}$. It is due to the fact that the valence band
$\Gamma_8$ and spin-orbit split band $\Gamma_7$ contribute to
the matrix element of spin-flip transitions with  opposite signs.
\section{{Acknowledgements}}
We thank D. Weiss and W. Wegscheider for many helpful discussions.
Financial support from the DFG, the RFFI, the Russian Ministry of Science and
the NATO linkage program is gratefully acknowledged.

%

\end{document}